# Nested Intervals Tree Encoding with Continued Fractions

VADIM TROPASHKO[1]
Oracle Corp.

> There is nothing like abstraction
> To take away your intuition
> *Shai Simonson*
> http://aduni.org/courses/discrete/

________________________________________________________________
We introduce a new variation of Tree Encoding with Nested Intervals, find connections with Materialized Path, and suggest a method for moving parts of the hierarchy.

Categories and Subject Descriptors: H.2 [**Database Management**]:
________________________________________________________________

## 1. INTRODUCTION

There are several SQL techniques to query graph structures, in general, and trees, in particular. They can be classified into 2 major categories:

- Hierarchical/recursive SQL extensions
- Tree Encodings

This article focuses upon Tree Encodings.

Tree encodings methods themselves can be split into 2 groups:

- Materialized Path
- Nested Sets

Materialized Path is nearly ubiquitous encoding, where each tree node is labeled with the path from the node to the root. UNIX global filenames is well known showcase for this idea. Materialized path could be either represented as character string of unique sibling identifiers (concatenated with some separator), or enveloped into user defined type (*Roy [2003]*).

Querying trees with Materialized Path technique doesn't appear especially elegant. It implies either string parsing, or leveraging complex data types that are realm of Object-Relational Databases. The alternative tree encoding - Nested Sets (*Celko [2000],[2004]*) labels each node with just a pair of integers. Ancestor-descendant relationship is reflected by subset relation between intervals of integers, which provides very intuitive base for hierarchical queries. A slight variation of Nested Sets is Dietz labeling with `<preorder#, postorder#>` pair of integers (*Li et al. [2001]*). The linear mapping

```
left = total#nodes – postorder# +1
right = 2*total#nodes – preorder#
```

translates Dietz schema into Nested Intervals with integer boundaries (also called Nested Sets with gaps).

---
[1] Authors' email address: Vadim.Tropashko@orcl.com.

Although Nested Sets elegant technique was certainly appealing to many database developers, it has 2 fundamental disadvantages:
- The encoding is volatile. In a word, roughly half of the tree nodes should be relabeled whenever a new node were inserted.
- Querying ranges is asymmetric from performance perspective. It is easy to answer if a point falls inside some interval, but it is hard to index a set of intervals that contain a given point. For Nested Sets this translates into a difficulty answering queries about node's ancestors.

*Tropashko [2003a]* introduced Nested Intervals that generalize Nested Sets. Since Nested Sets encoding with integers allows only finite gaps to insert new nodes, it is natural to use dense domain such as rational numbers. One particular encoding schema with Binary Rational Numbers was developed in the rest of the article, and was a subject of further improvements in the follow up articles. Binary Rational Encoding has many nice theoretical properties, and essentially is a numeric reflection of Materialized Path. It has, however, one significant flaw from practical perspective. Binary Fractions utilize domain of integer numbers rather uneconomically, so that numeric overflow prevents tree scaling to any significant size.

In general, Nested Intervals allow a certain freedom choosing particular encoding schema. *Tropashko [2004]* developed alternative encoding with Farey Fractions. It solved scalability problems, but it remained unclear how this new encoding is related to Materialized Path. Furthermore, a predictable question from developers' community was "How to relocate subtrees in this new schema?"

This article addresses both concerns. We expand our perspective and invoke some elementary math methods, including Continued Fractions, Greatest Common Divisor (GCD), Euclid Algorithm, 2x2 Matrix Algebra. This technique provides unexpected insight into connection of Farey Encoding with Materialized Path.

2. THE ENCODING

We label tree nodes with rational numbers `a/b` such that `a≥b≥1` and `GCD(a,b)=1`. Node with `a=4913` and `b=1594` would be used as our primary example through the entire article. Euclidean Algorithm maps the `4913/1594` node it into a sequence

```
4913 = 1594*3  + 131
1594 = 131 *12 + 22
131  = 22  *5  + 21
22   = 21  *1  + 1
21   = 1   *21 + 0
```

that we interpret as Materialized Path `3.12.5.1.21`. The opposite mapping – from Materialized Path to Rational Numbers – is implemented via Continued Fractions.

3. CONTINUED FRACTIONS

Simple Continued Fraction is a list of integers structurally arranged like this:

$$3 + \cfrac{1}{12 + \cfrac{1}{5 + \cfrac{1}{1 + \cfrac{1}{21}}}}$$

When converting Continued Fraction into Rational Number, we go through the steps of Euclidean Algorithm in the reverse order, so that in our example we would necessarily get

$$3 + \cfrac{1}{12 + \cfrac{1}{5 + \cfrac{1}{1 + \cfrac{1}{21}}}} = \frac{4913}{1594}$$

4. NESTED INTERVALS

Nested Intervals *Tropashko [2003a]* enjoy desirable Nested Sets properties and, therefore, it is important that continued fractions could be interpreted as nested intervals.

We map every Continued Fraction into a [semiopen] interval in 2 steps. First, we associate a Rational Function with every Continued Fraction as follows:

$$3 + \cfrac{1}{12 + \cfrac{1}{5 + \cfrac{1}{1 + \cfrac{1}{21 + \cfrac{1}{x}}}}} = \cfrac{4913\,x + 225}{1594\,x + 73}$$

Then, we assume $x \in [1, \infty)$. Substituting the boundary values for $x$ into the Rational Function for `3.12.5.1.21` we find that it ranges inside the `(4913/1594, 5138/1667]` semiopen interval.

Let's demonstrate that the intervals are nested. Indeed, if $x$ is allowed to be any number in the range $[1, \infty)$, then we could substitute $x$ for another Rational Function! Therefore, nesting Rational Functions derived from Continued Fractions corresponds to Materialized Path Concatenation. For example, concatenating paths `3.12` and `5.1.21` is nesting

$$x = 5 + \cfrac{1}{1 + \cfrac{1}{21 + \cfrac{1}{y}}}$$

inside of

$$3.12 = 3 + \cfrac{1}{12 + \cfrac{1}{x}}$$

which is a formal substitution of variable *x*.

Let's double check that the interval for `3.12.5.1.21` is indeed nested inside the interval corresponding to the path `3.12`. We have `[3+1/(12+1/1),3+1/12)` = `[40/13,37/12)` and `40/13 < 4913/1594 < 5138/1667 < 37/12`.

Rational function with linear numerator and denominator polynomials is called *Möbius transformation* and, therefore, we'll refer to the latest tree encoding as *Möbius encoding*. Let's summarize details of mapping among Rational numbers, Continued fractions,

Möbius encodings, and Intervals with Rational Boundaries that we have already established:

- Rational number → Materialized Path (by Euclid Algorithm)
- Materialized Path → Rational number (by simplifying Continued Fraction)
- Möbius encoding `(ax+b)/(cx+d)` ↔ Interval `(a/c, (a+b)/(c+d)]`
- Interval `(a/c, (a+b)/(c+d)]` → Rational number `a/c`
- Materialized Path → Möbius encoding (appending `1/x` to Continued Fraction and simplifying)

We'll use these encodings interchangeably. In the next section we'll show that Rational Number encodings for the parent and sibling are expressed in terms of original node's Möbius representation with astonishing simplicity.

## 5. PARENT AND NEXT SIBLING

***Lemma 1***. `225/73` is parent of `(4913x+225)/(1594x+73)`.

***Proof***. Assume that parent encoding in Möbius representation is `(ay+b)/(cy+d)`, where we changed free variable to `y`. Then, concatenating path `3.12.5.1` with `21` corresponds to nesting `y=21+1/x` inside `(ay+b)/(cy+d)`. By substitution we have `((21a+b)x+a)/((21c+d)x+c)` which, on the other hand, should be equal to `(4913x+225)/(1594x+73)`. Therefore, `a=225`, `c=73`.

***Lemma 2***. `(4913+225)/(1594+73)` is the next sibling of `(4913x+225)/(1594x+73)`.

***Proof***. As we already established in lemma 1, parent encoding is `(225y+b)/(73y+d)`. Also in lemma 1 we nested Möbius encoding `y=21+1/x` for node `21` inside its parent, and got resulting encoding `((21*225+b)x+225)/((21*72+d)x+73)`. If we were nested encoding `y=22+1/x` for node `22` instead, then, we would have got `((22*225+b)x+225)/((22*72+d)x+73)`. Given that `21*225+b=4913` and `21*72+d=1594` we immediately have `22*225+b=4913+225` and `22*72+d==1594+72`.

With the help of lemma 1 we expect the next sibling's Möbius encoding to be `((4913+225)x+225)/((1594+73)x+73)`

## 6. MAIN LEMMA

***Lemma 3***. Let `(ax+b)/(cx+d)` is Möbius encoding. Then, either `bc - ad = 1` or `ad - bc = 1`.

***Proof***. Induction by nesting level. Assume `bc = ad + 1`. Increasing level one more corresponds to nesting `x=n+1/y` inside the parent `(ax+b)/(cx+d)` for some integer n. New Möbius encoding is `(any+by+a)/(cny+dy+c)` and we only have to verify that `acn+ad=acn+bc+1`. The other case is symmetric.

Lemma 3 has at least two interpretations. Möbius transformation `(ax+b)/(cx+d)` parallels algebra of 2x2 matrices

$$\begin{bmatrix} a & b \\ c & d \end{bmatrix}$$

where composition of Möbius transformations corresponds to matrix multiplication. Lemma 3 constraints matrix algebra to those matrices that have determinants -1 or 1 only. Without going into too much detail let's represent path `3.12.5.1.21` as matrix product

$$\begin{bmatrix} 3 & 1 \\ 1 & 0 \end{bmatrix} \cdot \begin{bmatrix} 12 & 1 \\ 1 & 0 \end{bmatrix} \cdot \begin{bmatrix} 5 & 1 \\ 1 & 0 \end{bmatrix} \cdot \begin{bmatrix} 1 & 1 \\ 1 & 0 \end{bmatrix} \cdot \begin{bmatrix} 21 & 1 \\ 1 & 0 \end{bmatrix} = \begin{bmatrix} 4913 & 225 \\ 1594 & 73 \end{bmatrix}$$

where each one node path primitive matrix has determinant -1 and, therefore, it's obvious that the multiplication result should have determinant 1 or -1.

We provide another interpretation of Lemma 3 in terms of GCD in the next section.

6. EXTENDED EUCLIDEAN ALGORITHM

Extended version of the Euclidean algorithm calculates three numbers `GCD(a,b)`, `x` and `y` which meet the following identity

```
ax-by=GCD(a,b)
```

Since `GCD(a,b)=1` for our rational encodings, then the above identity coincides with the one from Lemma 3. Therefore, we could use Extended Euclidean algorithm to calculate Möbius encoding if we know rational representation. Let's demonstrate this for the familiar node `4913/1594`

```
4913-1594*3=                                         = 4913*1 -1594*3  = 131
1594-131*12= 1594           -(4913*1-1594*3)   *12= 1594*37-4913*12 = 22
131 -22*5  = (4913*1-1594*3)  -(1594*37-4913*12) *5= 4913*61-1594*188= 21
22  -21*1  = (1594*37-4913*12)-(4913*61-1594*188)*1= 1594*225-4913*73= 1
```

As expected, the result `225/73` is the parent encoding. This algorithm presents significant speed improvement compared to naïve loop iterating through all numbers from `1` to `denom` when finding next Farey fraction in *Tropashko [2004]* [2]

## 7. RELOCATING SUBTREES

Consider subtree rooted at the node `3.12`. When relocating all the descendants of `3.12` we'll apply a set operation like in *Tropashko [2003a]*, but for clarity we focus on single node `3.12.5.1.21` only. We want to "detach" this node from its ancestor, first. Speaking path language, we represent `3.12.5.1.21` as a concatenation of path fragments `3.12` and `5.1.21` so that `5.1.21` can be later reattached to the other parent. Speaking matrix algebra language from section 5, we multiply matrix corresponding to `3.12` to unknown matrix and the result have to be equal to matrix corresponding to `3.12.5.1.21`. Therefore, we have to solve matrix equation

$$\begin{bmatrix} 37 & 3 \\ 12 & 1 \end{bmatrix} \cdot \begin{bmatrix} x11 & x12 \\ x21 & x22 \end{bmatrix} = \begin{bmatrix} 4913 & 225 \\ 1594 & 73 \end{bmatrix}$$

which reduces to a system of 4 linear equations. The solution matrix is

$$\begin{bmatrix} 131 & 6 \\ 22 & 1 \end{bmatrix}$$

We could have calculated this matrix directly, because we know the path fragment `5.1.21` it corresponds to

$$\begin{bmatrix} 5 & 1 \\ 1 & 0 \end{bmatrix} \cdot \begin{bmatrix} 1 & 1 \\ 1 & 0 \end{bmatrix} \cdot \begin{bmatrix} 21 & 1 \\ 1 & 0 \end{bmatrix} = \begin{bmatrix} 131 & 6 \\ 22 & 1 \end{bmatrix}$$

but the idea is that we cold do calculations in any of the encodings that we introduced in this article without being forced to translate to Materialized path.

Finally, let's attach this path fragment the other part of hierarchy, say under the path `4.7`. In matrix language the matrix corresponding to `4.7`

$$\begin{bmatrix} 4 & 1 \\ 1 & 0 \end{bmatrix} \cdot \begin{bmatrix} 7 & 1 \\ 1 & 0 \end{bmatrix} = \begin{bmatrix} 29 & 4 \\ 7 & 1 \end{bmatrix}$$

should be [right] multiplied by the matrix corresponding to the moving path fragment `5.1.21` that we found previously

---

[2] Parent node encoding in *Tropashko [2004]* is expressed in terms of next Farey fraction

$$\begin{bmatrix} 29 & 4 \\ 7 & 1 \end{bmatrix} \cdot \begin{bmatrix} 131 & 6 \\ 22 & 1 \end{bmatrix} = \begin{bmatrix} 3887 & 178 \\ 939 & 43 \end{bmatrix}$$

which concludes our single node relocation.

## CONCLUSION

In this article we focused on tree encodings with rational numbers that are greater or equal to 1. Farey Fractions tree encoding in *Tropashko [2004]* used rational numbers that are less or equal to 1. This choice, was incidental, because Farey sequence is symmetrical with respect to multiplicative inverse operation. Likewise, in this article we could have studied continued fractions and rational functions of the kind:

$$\cfrac{1}{3 + \cfrac{1}{12 + \cfrac{1}{5 + \cfrac{1}{1 + \cfrac{1}{20 + x}}}}}$$

where both `x` and the value rational function is between `0` and `1`. Some our results would have changed to multiplicative inverse. The fundamental identity `ad - bc = 1`, however, holds for two adjacent fractions `a/c` and `b/d` in all these encoding schema variations.

## REFERENCES

CELKO, J. 2000. Trees in SQL. http://www.intelligententerprise.com/001020/celko.shtml
CELKO, J. 2004. *Joe Celko's Trees and Hierarchies in SQL for Smarties*. Morgan Kaufmann.
LI, Q., MOON, B. 2001. Indexing and Querying XML Data for Regular Path Expressions. *The VLDB Journal*. (http://citeseer.ist.psu.edu/li01indexing.html)
ROY, J. 2003. Using the Node Data Type to Solve Problems with Hierarchies in DB2 Universal Database http://www-106.ibm.com/developerworks/db2/library/techarticle/0302roy/0302roy.html
TROPASHKO, V. 2003a. Trees in SQL: Nested Sets and Materialized Path. http://www.dbazine.com/tropashko4.shtml
TROPASHKO, V. 2003b. Relocating Subtrees in Nested Intervals Model. http://www.dbazine.com/tropashko5.shtml
TROPASHKO, V. 2004. Nested Intervals with Farey Fractions. http://arxiv.org/html/cs.DB/0401014